\documentclass{article}
\usepackage{graphicx}
\usepackage{amssymb}
\usepackage{amsmath}
\usepackage{mathtools}
\usepackage{float}
\usepackage[numbers]{natbib}
\usepackage{url}
\usepackage{gensymb}

\usepackage[
    left=2.5cm,
    right=2.5cm,
    top=2.5cm,
    bottom=2.5cm,
]{geometry}

\title{ Experimental Insights into the Limiting Mechanism of Vacancy Transport in Sodium Metal Anodes for Solid State Batteries}
\begin{document}

\maketitle

\author{Ansgar Lowack*}
\author{Rafael Anton}
\author{Bingchen Xue}
\author{Kristian Nikolowski}
\author{Cornelius Dirksen}
\author{Mareike Partsch}
\author{Alexander Michaelis}

\subsubsection*{Affiliations}
A. Lowack, Prof. Dr. A. Michaelis\\
Institute of Material Science\\
Dresden University of Technology\\
Helmholtzstraße 7, 01069 Dresden, Germany
Email Address: ansgar.lowack@tu-dresden.de

A. Lowack, R. Anton, B. Xue, Dr. K. Nikolowski, Dr. C. Dirksen, Dr. M. Partsch, Prof. Dr. A. Michaelis\\
Fraunhofer Institute for Ceramic Technologies and Systems (IKTS)\\
Winterbergstraße 28, 01277 Dresden, Germany

\subsubsection*{Keywords}
Sodium metal electrode, Solid state battery, Void formation, Sodium solid electrolyte, Ceramic ion conductor, Interfacial electrochemistry, Sodium vacancy transport

\begin{abstract}

\noindent Ceramic solid-state batteries with sodium (Na) metal electrodes promise enhanced safety and energy density compared to contemporary secondary batteries. However, the critical delamination of the Na metal electrode during discharge - when vacancies accumulate at the Na/ceramic interface - remains to be understood and avoided. The study investigates the underlying mechanism by applying a linear current ramp between two Na metal electrodes separated by a ceramic solid electrolyte to provoke vacancy buildup. Above a critical current density $j_\mathrm{crit}$ the anode voltage no longer increases linearly but in an exponential fashion. Arrhenius analysis of $j_\mathrm{crit}(T)$ for the three solid electrolytes $\mathrm{Na_{1.9}Al_{10.67}Li_{0.33}O_{17}}$, $\mathrm{Na_{3.4}Zr_2Si_{2.4}P_{0.6}O_{12}}$, and $\mathrm{Na_5SmSi_4O_{12}}$ yields an activation energy $E_\mathrm{A}$ of $0.13$ to $0.15\,\mathrm{eV}$. This exceeds the activation energy of $0.053\,\mathrm{eV}$ for the diffusive vacancy migration in bulk Na significantly. Further, $E_\mathrm{A}$ is insensitive to anode microstructure variation. Both observations rule out bulk diffusion as the transport bottleneck. A thin tin-sodium alloy interlayer lowers $E_\mathrm{A}$ to $(0.10\pm0.01)\,\mathrm{eV}$, implicating interfacial thermodynamics as rate-limiting. Sodiophilic, Na-conducting interlayers and low-tension interfaces emerge as key pathways to stable, high-rate Na-SSBs at practical stack pressures.

\end{abstract}

\section{Introduction}
\label{introduction}
As the global demand for advanced battery cells grows, ceramic sodium (Na) solid-state batteries (SSBs) have entered the technological discussion. There is good reason to believe that these devices will be inherently safer than the notoriously flammable lithium (Li) -ion batteries, since flammable liquids are replaced with ceramic solid electrolytes (SE) \cite{Janek.2023}. While Li-based SSBs share this safety advantage, Na is particularly promising due to the availability of lower-density, higher-stability, and higher-conductivity SEs compared to Li analogues \cite{Radjendirane.2024, Wang.2020b, Sazvar.2025}, such as $\mathrm{Na_{1.9}Al_{10.67}Li_{0.33}O_{17}}$ \cite{Fertig.2022, Fertig.2022b} ("$\mathrm{Na-\beta-alumina}$"),\\ $\mathrm{Na_{3.4}Zr_2Si_{2.4}P_{0.6}O_{12}}$ \cite{Ma.2020, Liu.2025, Liu.2025b, Ma.2019b} ("NASICON") and $\mathrm{Na_5SmSi_4O_{12}}$ \cite{Schilm.2022, Anton.2025, Michalak.2024} SEs. Furthermore, SSBs may enable safe and reversible cycling of Na metal electrodes at low interface resistance \cite{Ortmann.2023, Ma.2022, Huttl.2022}, eliminating the need for graphite or hard carbon commonly used in liquid-electrolyte Li- or Na-ion secondary batteries and opening pathways to higher energy density \cite{Lowack.2025b}. However, despite these promises, no commercial Na‑SSBs with ceramic solid electrolyte and sodium metal electrode that rival conventional Li‑ion battery performance are available, as fundamental challenges remain to be understood and addressed.

One remaining key issue is the accumulation of atomic vacancies at the interface between the metal electrode and the SE during cell discharge, leading to void formation and anode delamination. The consequences of this process include a sharp, irreversible rise in interfacial impedance, significant capacity fading, and an increased susceptibility to dendrite-induced short-circuiting during subsequent charging cycles. This phenomenon had been under investigation three decades ago in a different context by Schmalzried, Janek et al. \cite{Janek.1997, Schmalzried.1998} and has gained recent attention for both Li and Na metal electrodes in the context of SSBs, with insightful works on the topic \cite{Lang.2025,Lu.2022, Wang.2025, Cheng.2025, Tang.2025}. However, the underlying mechanisms remain subject of debate as the physics are a complex interplay between diffusive transport modes \cite{Yoon.2023}, interfacial effects \cite{Seymour.2021}, plastic deformation \cite{Sargent.1984} and void growth mechanisms \cite{Schmalzried.1998, Schroder.2006}, among other effects. 

Prior experimental studies have focused on the phenomenology and the imaging of void morphologies across different current regimes to infer growth mechanisms \cite{Lang.2025, Lu.2022, Ortmann.2023, Fuchs.2024, Wang.2025}, but there is a lack of experimental work that systematically probes the initial transport limitations of interfacial vacancies that trigger void nucleation. This study addresses this gap in four steps: (i) An intuitive physical model is sketched for plausible mechanisms limiting the transport of interfacial vacancies into the electrode bulk in the absence of significant stress driven sodium creep. This allows to distinguish the mechanisms by their activation energies. (ii) It is demonstrated how interfacial void accumulation can be electrochemically induced and measured in model cells with  Na metal anode (electrode where Na is oxidized), Na metal cathode (electrode where Na is reduced), a ceramic SE separator and a Na metal reference electrode. This allows the  definition of a critical current density of Na oxidation (i.e. related to the discharge of a SSB). Here, the distinction to cell failure during Na reduction (i.e. related to the charging of a SSB) must be emphasized. The latter can be attributed to the penetration of Na dendrites through the ceramic SE which is commonly quantified by a critical current density of Na reduction. While both failure modes influence each other due to locally high (dendrites) or locally low (voids) currents across the interface, when cycling a SSB, the effects have unrelated physical origins. Cell failure during Na reduction and dendrite formation are not content of the presented study. For further reading on this topic, the existing literature is referred \cite{VIRKAR.1979, Ning.2023, Geng.2023, Liu.2020, Yu.2021} including our own \cite{Lowack.2025}. (iii) The study measures the temperature dependence of this critical current density of oxidation at neglectable external stack pressure (see Experimental Section) and compares it with hypothetical mechanisms. To gain insight, SE chemistry and anode design are varied: cells with $\mathrm{Na_{1.9}Al_{10.67}Li_{0.33}O_{17}}$, $\mathrm{Na_{3.4}Zr_2Si_{2.4}P_{0.6}O_{12}}$ and $\mathrm{Na_5SmSi_4O_{12}}$ SE are assembled, and the Na metal anode is altered in microstructure and by introducing an alloyed Sn-Na interlayer to assess impacts on vacancy accumulation. (iv) From these experiments, the mechanism that limits vacancy transport and thus constitutes the bottleneck for engineering robust Na-SSBs is identified.

\section{Results and Discussion}
\label{results_discussion}
\subsection{Theory of sodium vacancy transport}
Key aspects of vacancy transport in Na metal anodes for SSBs are modeled to explain experiment design. The works of Seymour, Aguadero, Yoon, et al. are recommended for further theoretical discussion \cite{Seymour.2021, Yoon.2023}.

\begin{figure}[h]
    \centering
    \includegraphics[width=0.8\linewidth]{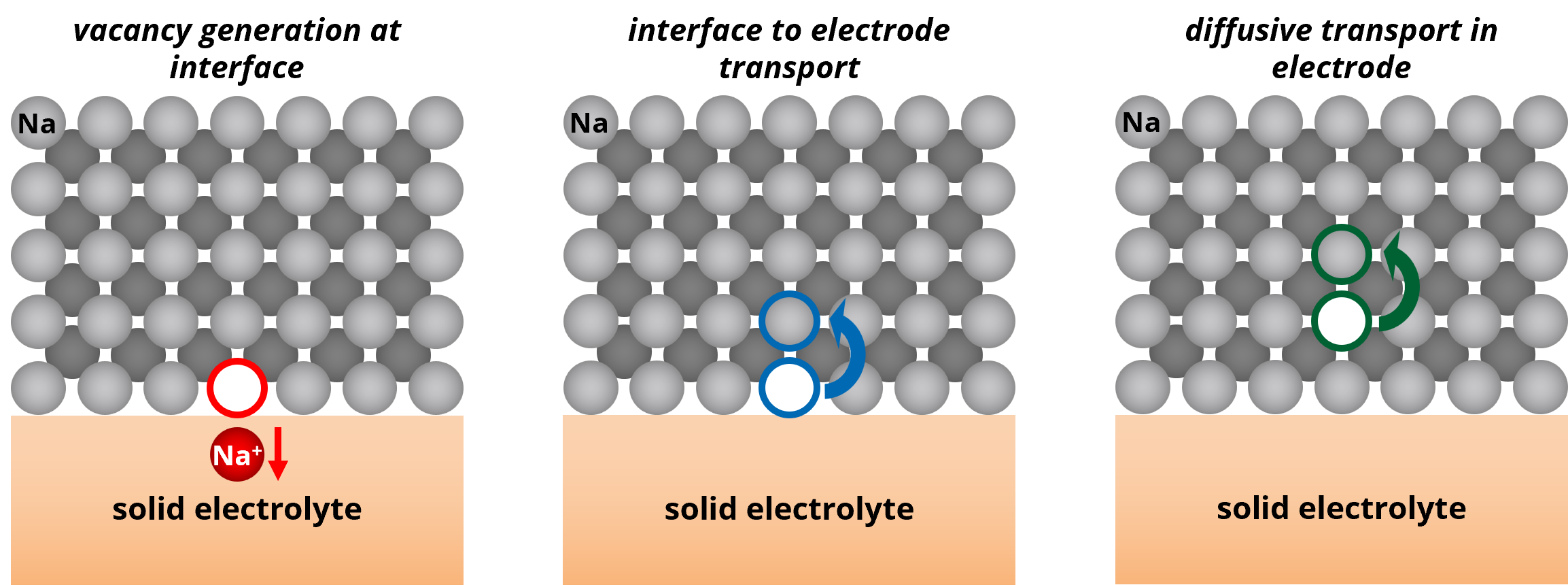}
    \caption{Schematic depiction of the simplified vacancy-transport mechanism. A sodium vacancy is generated at the interface by oxidation (red), it is thermally excited into the electrode bulk (blue), and diffusively transported within the electrode bulk (green).}
    \label{fig:Mechanism}
\end{figure}

The vacancy transport is decomposed into three steps: vacancy generation at the Na/SE interface, transfer from the interface into the electrode, and vacancy diffusion within the electrode, as illustrated in \textbf{Figure \ref{fig:Mechanism}}. The creep of the Na – its mechanical relaxation – is considered negligible at the low stack pressures utilized in this study (see Experimental Section).

\medskip
\noindent\textbf{Interfacial generation (red in Figure \ref{fig:Mechanism}).}
The interfacial vacancy generation rate $\dot{N}_\mathrm{generation}$ is proportional to the oxidation current density $j$:
\begin{equation}
\label{eq:Generation}
    \dot{N}_\mathrm{generation}=\kappa j,
\end{equation}
with a constant $\kappa$. Here and throughout, $j$ is defined per cell cross-sectional area rather than the true contact area between SE and electrode.

\medskip
\noindent\textbf{Diffusive transport in electrode (green in Figure \ref{fig:Mechanism}).}
Vacancy transport in the Na metal electrode is governed by Fick’s laws of diffusion. The relevant mechanism is \emph{vacancy} diffusion, not Na self-diffusion. The activation energy of vacancy transport is the migration energy $E_\mathrm{m}$ for a Na atom hopping into a neighboring vacancy. In the bcc Na lattice (lat), this value is well known as $E^\mathrm{lat}_\mathrm{m}=(0.053\pm0.001)\,\mathrm{eV}$ from calculations and experiments \cite{Ma.2019, VSchott.2000, Ullmaier1991}, which is significantly smaller than the self-diffusion activation energy, $E^\mathrm{lat}_\mathrm{SD}=E^\mathrm{lat}_\mathrm{m}+E^\mathrm{lat}_\mathrm{f}=(0.39\pm0.01)\,\mathrm{eV}$, with the vacancy formation energy $E^\mathrm{lat}_\mathrm{f}=(0.34\pm0.01)\,\mathrm{eV}$ \cite{Ma.2019, Ullmaier1991}. In polycrystalline Na (as used below), vacancy diffusion at grain boundaries differs from lattice diffusion. This phenomenon is treated in detail for Li by Yoon et al. \cite{Yoon.2023}. To reduce the problem to one dimension, the temperature-dependent effective vacancy diffusion coefficient $D^\mathrm{eff}(T)$ is introduced via Hart’s equation:
\begin{equation}
\label{eq:Deff}
    D^\mathrm{eff}(T)=fD_0^\mathrm{gb}\exp\!\left(-\frac{E_\mathrm{m}^\mathrm{gb}}{k_\mathrm{B}T}\right)+(1-f)D_0^\mathrm{lat}\exp\!\left(-\frac{E_\mathrm{m}^\mathrm{lat}}{k_\mathrm{B}T}\right),
\end{equation}
where $D_0^\mathrm{gb}$ and $D_0^\mathrm{lat}$ are prefactors for grain-boundary (gb) and lattice (lat) vacancy diffusion, $E^\mathrm{gb}_\mathrm{m}$ and $E^\mathrm{lat}_\mathrm{m}$ are the corresponding migration energies, and $f$ is a geometry parameter capturing grain-boundary shape, spacing, and thickness. In bcc metals, the disordered structure at grain boundaries reduces coordination and increases local free volume, lowering the migration barrier ($E^\mathrm{gb}_\mathrm{m}<E^\mathrm{lat}_\mathrm{m}$). In one dimension, the vacancy flux $\xi$ away from the interface obeys Fick’s first law:
\begin{equation}
\label{eq:Fick1}
    \xi=-D^\mathrm{eff}(T)\frac{\partial c(x)}{\partial x},
\end{equation}
where $c(x)$ is the vacancy concentration in the Na electrode. Combining equation \ref{eq:Generation} and \ref{eq:Fick1} yields a first critical-current condition (steady state requires that interfacial generation does not exceed the diffusive flux into the electrode):
\begin{equation}
\label{eq:crit1}
    j\leq \frac{|\xi|}{\kappa}=\frac{D^\mathrm{eff}(T)}{\kappa}\left|\frac{\partial c(x)}{\partial x}\right|.
\end{equation}
If oxidation proceeds at a current density that violates equation \ref{eq:crit1}, vacancies accumulate at the interface and voids form. Because the equilibrium vacancy concentration in Na metal is negligible at the temperatures used here (e.g., $\exp(-E^\mathrm{lat}_\mathrm{f}/k_\mathrm{B}T)\sim 10^{-6}$ at $T=90\,\degree \mathrm{C}$) relative to concentrations that destabilize the lattice and nucleate voids (e.g., $\sim 10^{-2}$), its temperature dependence is neglected. Consequently, the temperature dependence of the first criterion follows equation \ref{eq:Deff}. If $j$ becomes critical due to bulk diffusion limitations, the measured activation energy $E_\mathrm{A}$ should satisfy
\begin{equation}
\label{eq:crit1_activation_energy}
   E^\mathrm{gb}_\mathrm{m}<E_\mathrm{A}<E^\mathrm{lat}_\mathrm{m}.
\end{equation}

\medskip
\noindent\textbf{Interface to electrode transport (blue in Figure \ref{fig:Mechanism}).}
Consider the vacancy formation energy $E^{\mathrm{int}}_\mathrm{f}$ at the Na/SE interface and $E_{\mathrm{f}}^\mathrm{Na}\approx E^\mathrm{lat}_\mathrm{f}=(0.34\pm0.01)\,\mathrm{eV}$
in bulk Na \cite{Ma.2019, Ullmaier1991}. Following thermodynamical arguments, $E^{\mathrm{int}}_\mathrm{f}$ depends on interfacial bonding and correlates with the Na/SE interfacial tension. For sodiophilic (low-tension) interfaces, the average Na–SE bonding is strong relative to Na–Na bonding, yielding $E^{\mathrm{int}}_\mathrm{f}>E_{\mathrm{f}}^{\mathrm{Na}}$. For sodiophobic (high-tension) interfaces, Na–Na bonding dominates, so $E^{\mathrm{int}}_\mathrm{f}<E_{\mathrm{f}}^{\mathrm{Na}}$. Ceramic SEs are typically sodiophobic, as evidenced by large contact angles to molten Na (see Supporting Information S1). At room temperature, high interfacial resistances between Na and ceramic SEs are commonly attributed to current constriction arising from poor wetting \cite{Eckhardt.2022, Ortmann.2023}. Consequently, an energy barrier $E_{\mathrm{f}}^\mathrm{Na}-E^{\mathrm{int}}_\mathrm{f}$ arises between vacancies at the interface and in the bulk of the Na electrode. A similar and more detailed argumentation of this effect can be found in \cite{Seymour.2021}. Vacancies which are generated at the interface must be thermally excited into the electrode bulk before they can be transported via vacancy diffusion in the Na metal.
Following the argumentation, interfacial vacancy transfer is modeled as a classical two-level system with ground state $E^{\mathrm{int}}_\mathrm{f}$ (vacancy energy at the interface) and excited state $E_{\mathrm{f}}^\mathrm{Na}$ (vacancy energy in the Na electrode). Only vacancies thermally excited to $E_{\mathrm{f}}^\mathrm{Na}$ (i.e. activated from the interface into the bulk of the Na electrode) can diffuse away from the interface. Let $k_\uparrow$ be the excitation rate and $k_\downarrow$ the relaxation rate. Detailed balance gives
\begin{equation}
    \frac{k_\uparrow}{k_\downarrow}=\exp\!\left(-\frac{E_{\mathrm{f}}^\mathrm{Na}-E^{\mathrm{int}}_\mathrm{f}}{k_\mathrm{B}T}\right).
\end{equation}
Assuming $k_\downarrow$ is only weakly temperature dependent, it is approximated as a constant attempt frequency $k_\downarrow\approx\nu$, yielding
\begin{equation}
\label{eq:RateExcitation}
    k_\uparrow =\nu\exp\!\left(-\frac{E_{\mathrm{f}}^\mathrm{Na}-E^{\mathrm{int}}_\mathrm{f}}{k_\mathrm{B}T}\right).
\end{equation}
Equation \ref{eq:Generation} to equation \ref{eq:RateExcitation} can be used to derive differential equations governing the time evolution of the number of vacancies at the interface for a given current density $j(t)$. However, since only the critical current density at which the interfacial vacancy number surpasses some critical value $N_\mathrm{crit}$ (and voids will start forming) is of interest, this is not needed here. It is sufficient to state a second critical current condition above which vacancies will accumulate at the interface and voids will form:
\begin{equation}
\label{eq:crit2}
    j\leq \frac{k_\uparrow N_\mathrm{crit}}{\kappa}
    =\frac{N_\mathrm{crit}\,\nu}{\kappa}\exp\!\left(-\frac{E_{\mathrm{f}}^\mathrm{Na}-E^{\mathrm{int}}_\mathrm{f}}{k_\mathrm{B}T}\right).
\end{equation}
Hence, if $j$ is limited by interfacial transport, $j_\mathrm{crit}(T)$ should obey Arrhenius behavior with activation energy
\begin{equation}
\label{eq:crit2_activation_energy}
    E_\mathrm{A}\approx E_{\mathrm{f}}^\mathrm{Na}-E^{\mathrm{int}}_\mathrm{f}.
\end{equation}

\medskip
\noindent\textbf{Theoretical summary.}
The Na-current $j$ will become critical due to interfacial vacancy accumulation when either current condition (equation \ref{eq:crit1} or \ref{eq:crit2}) is exceeded. Both arise from thermally activated processes with distinct activation energies $E_\mathrm{A}$. If vacancy transport in the Na electrode is rate-limiting (equation \ref{eq:crit1}), $E_\mathrm{A}$ is the vacancy migration energy in Na metal (equation \ref{eq:crit1_activation_energy}). If vacancy transport at the Na/SE interface is rate-limiting, $E_\mathrm{A}$ corresponds to the interfacial vacancy-excitation barrier (equation \ref{eq:crit2_activation_energy}) which must be overcome to thermally excite vacancies from the interface into the Na electrode.\\
The experimental objective is to extract $E_\mathrm{A}$ from measurements of $j_\mathrm{crit}(T)$ and thereby identify the limiting mechanism by comparison with equation \ref{eq:crit1_activation_energy} and \ref{eq:crit2_activation_energy}.

\subsection{Principle of the Measurement}
\begin{figure}[h]
    \centering
    \includegraphics[width=0.8\linewidth]{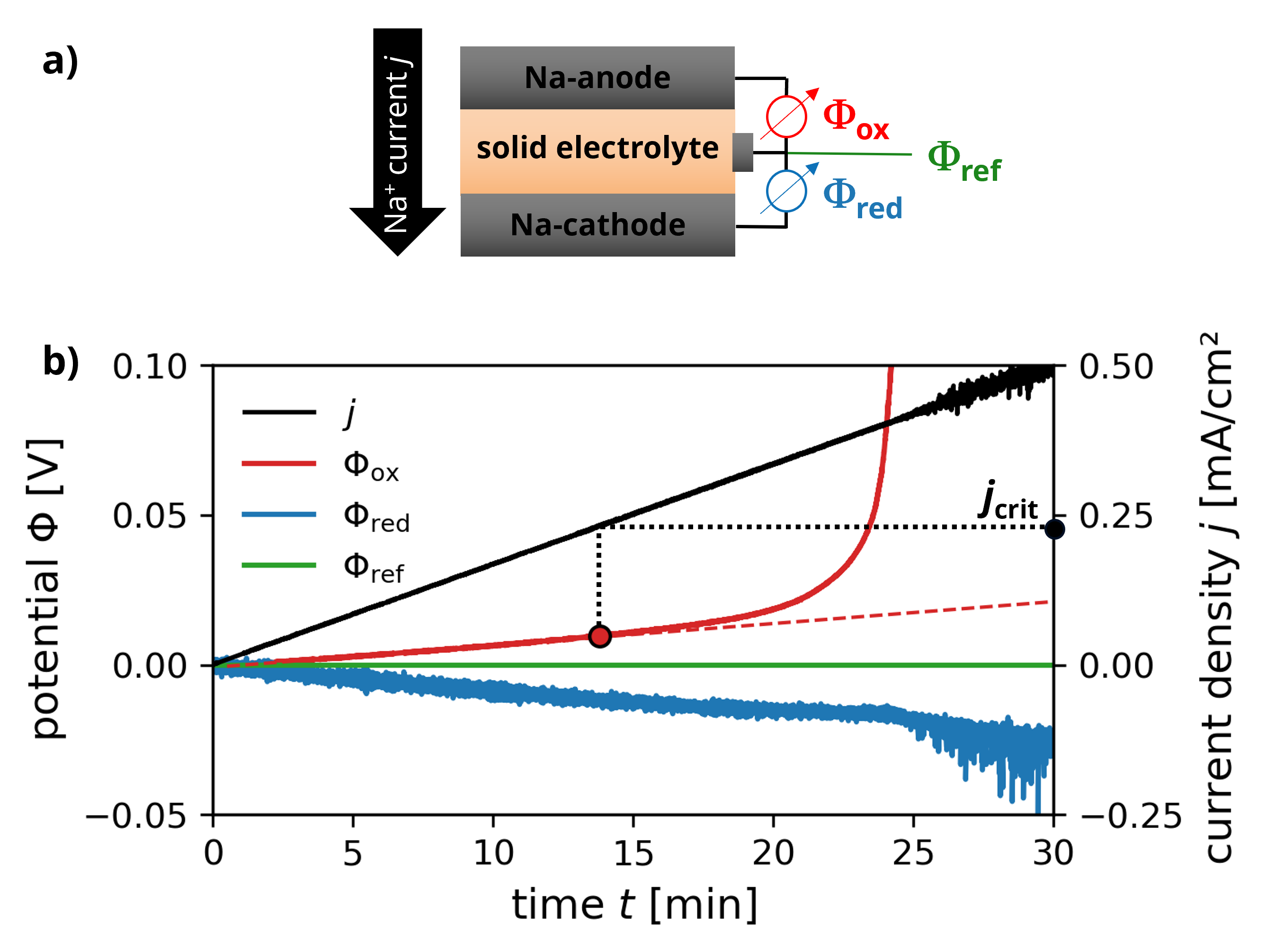}
    \caption{a) Schematic depiction of the three-electrode setup to measure void accumulation. b) Representative measurement of critical void accumulation at $I_\mathrm{crit}$ and $23\,\degree\mathrm{C}$.}
    \label{fig:Reference_Electrode}
\end{figure}

To quantify interfacial vacancy accumulation, symmetric cells with $\mathrm{Na_{3.4}Zr_2Si_{2.4}P_{0.6}O_{12}}$ SEs, a Na metal anode, a Na metal cathode and a Na metal reference electrode are assembled, as schematically depicted in \textbf{Figure \ref{fig:Reference_Electrode}}a. In this three-electrode configuration, a linearly increasing current density $j(t)$ with slope $1\,\mathrm{mA\,cm^{-2}\,h^{-1}}$ is imposed. Electrode potentials recorded during the ramp are shown in Figure \ref{fig:Reference_Electrode}b. The reference potential $\Phi_\mathrm{ref}$ is set to $0\,\mathrm{V}$. The anode potential $\Phi_\mathrm{ox}(j)$ (Na oxidation; red) becomes nonlinear once the current exceeds the critical value $j_\mathrm{crit}$. The value $j_\mathrm{crit}$ is defined as the point where the relative deviation of $\Phi_\mathrm{ox}(j)$ from a linear fit (dashed red line, details in the Experimental Section) first exceeds $5\,\%$. The cathode potential $\Phi_\mathrm{red}(j)$ (Na reduction; blue) remains approximately proportional to $j$, even beyond $j_\mathrm{crit}$. Note that beyond $t\approx 24\,\mathrm{min}$, rapid loss of anode contact causes the cell resistance to increase faster than the potentiostat can respond to, preventing accurate control of $j$ and measurement of $\Phi_\mathrm{red}$; the apparent noise is a measurement artifact.

The SE exhibits linear ohmic behavior over the probed voltage range (following the Nernst–Einstein relation), and Butler–Volmer nonlinearity at the Na/$\mathrm{Na_{3.4}Zr_2Si_{2.4}P_{0.6}O_{12}}$ interface cannot explain the observed rise in $U$ (as it would approximate an inverse hyperbolic sine function). Consequently, the observed nonlinearity in $\Phi_\mathrm{ox}(j)$ is attributed to loss of interfacial contact area between anode and SE, i.e. an increase in interfacial constriction resistance that originates from the formation of interfacial voids \cite{Eckhardt.2022}. 

It must be emphasized that void formation is not synonymous but a consequence of the interfacial vacancy accumulation. Hence, a proper definition of $j_\mathrm{crit}$ in the sense of the two critical conditions (equation \ref{eq:crit1} and \ref{eq:crit2}) requires further consideration: Firstly, the delay between the onset of vacancy accumulation and the emergence of void growth must be estimated. A characteristic time $\tau$ for vacancy accumulation to nucleate void growth can be obtained by treating the process as two-dimensional vacancy diffusion along the interface over a diffusion length $l$:
\begin{equation}
    \label{eq:time_scale}
    \tau=\frac{l^{2}}{D_\mathrm{int}(T)}=\frac{l^{2}}{D_0^\mathrm{int}\exp\!\left(-\frac{E_{\mathrm{m}}^\mathrm{int}}{k_\mathrm{B}T}\right)},
\end{equation}
where $D_0^\mathrm{int}$ is the pre-exponential factor and $E_{\mathrm{m}}^\mathrm{int}$ the migration barrier of interfacial vacancy diffusion. Owing to the sodiophobic nature of ceramic SEs, $D_0^\mathrm{int}$ is expected to be comparable to vacancy diffusion on Na surfaces and thus of similar magnitude to low–migration-energy metal lattices \cite{Angsten.2014}, i.e., $D_0^\mathrm{int}\approx 10^{-6}\,\mathrm{m^{2}\,s^{-1}}$. The reduced coordination of interfacial Na relative to the bulk suggests $E_{\mathrm{m}}^\mathrm{int}\leq E^\mathrm{lat}_\mathrm{m}=(0.053\pm0.001)\,\mathrm{eV}$ which also means any two dimensional interfacial diffusion effects on vacancy accumulation cannot increase the measured activation energy. The relevant diffusion length is likely on the order of magnitude of the interfacial roughness, which is governed by the mean particle diameter ($\sim 2\,\mathrm{\mu m}$), as pellets were not polished to retain low initial contact resistances \cite{Ortmann.2023, Huttl.2022, Lowack.2025c}. A deliberately generous bound $l\leq 100\,\mathrm{\mu m}$ therefore applies. Notably, these values are much smaller than the SE thickness. The cathode interface will not influence the diffusion at the anodic interface. Inserting these values into equation \ref{eq:time_scale} yields $\tau < 1\,\mathrm{s}$ across the temperature range considered. On the time scale of the experiments (minutes), void nucleation following vacancy accumulation is thus effectively instantaneous. Secondly, the time required for voids, once nucleated, to grow to a size that measurably increases the constriction resistance must be discussed. A uniform coverage of nanoscale voids is already sufficient, as evidenced by the large constriction resistances observed between polished SEs and Na electrodes \cite{Eckhardt.2022, Ortmann.2023, Lowack.2025c}. A conservative upper bound is obtained by assuming that the removal of $\sim 10$ atomic Na layers across the \textit{entire} electrode after the onset of vacancy accumulation suffices to form significant voids; this corresponds to a charge of $0.4\,\mathrm{\mu Ah\,cm^{-2}}$. Under the measurement conditions, this charge translates into an uncertainty of $\Delta j_\mathrm{crit}\leq 0.004\,\mathrm{mA\,cm^{-2}}$ (find details in the Supporting Information S2), which is negligible.\\
\medskip
What must be mentioned however, is that due to the roughness and the polycrystallinity of the samples, current density on the interface is not homogeneous. Statistically, there must be small subareas on the interface, where vacancies start accumulating significantly below $j_\mathrm{crit}$ in Figure \ref{fig:Reference_Electrode}, due to local current focusing. This would quickly result in local void growth by the argumentation of equation \ref{eq:time_scale}. However, due to the initially microscopic size of this void, it would need to grow significantly in size until it alone has enough impact on interfacial resistance to be detectable by the measurement. While such local current criticality would follow the same activation energy, as the mechanistic difference is purely geometric, it is a three dimensional effect which is not described by the one dimensional model given above.  For the scope of this study, it is important that the current is ramped up quickly enough, so that this highly scattering local accumulation of voids does not dominate cell failure. Instead, voids must accumulate quickly enough on a sufficiently large portion of the interface to be detected with high reproducibility to accurately determine the activation energy. As will be seen in the following, this is the case at $1\,\mathrm{mAch^{-2}h^{-1}}$ as scattering is sufficiently low. However, consequently it is emphasized that determined numerical values of $j_\mathrm{crit}$ have little universal relevance, in contrast to the measured activation energies.\\
\medskip
To summarize the measurement principle: The emergence of nonlinear resistance in Na/SE/Na cells is traced to anode delamination. A relative nonlinearity of $5\,\%$ is defined as the critical current density of oxidation $j_\mathrm{crit}$ which should approximate the onset of vacancy accumulation with sufficient accuracy, given a ramping current density of $1\,\mathrm{mA\,cm^{-2}\,h^{-1}}$.

\subsection{Measuring vacancy accumulation}
To identify which condition drives $j$ to criticality, the temperature dependence of $j_\mathrm{crit}$ is measured. The value of $j_\mathrm{crit}$ is determined at $-30$, $0$, $30$, $60$, and $90\,\degree \mathrm{C}$ by evaluating the current at which the voltage deviates by more than $5\,\%$ from linear behavior. Figure \ref{fig:Reference_Electrode}b shows that the nonlinear voltage response originates at the anode. As the other two interfaces remain unchanged, no reference electrode is used during the measurement. Only the overall cell voltage $U=\Phi_\mathrm{ox}-\Phi_\mathrm{red}$ required to drive the $\mathrm{Na}^+$ current from anode to cathode is recorded as a function of $j$.

\begin{figure}[h]
    \centering
    \includegraphics[width=0.6\linewidth]{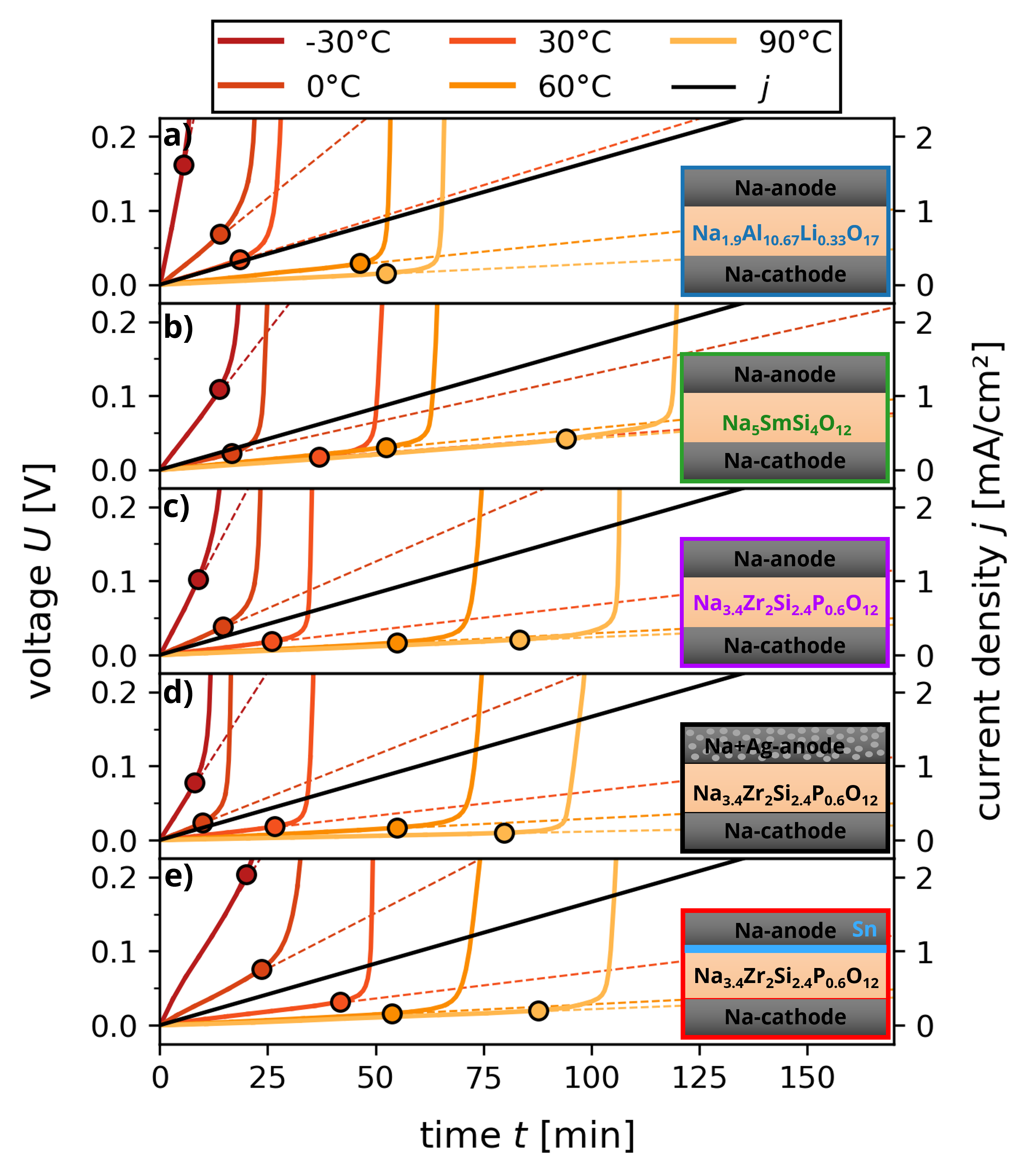}
    \caption{Determination of the critical current density $j_\mathrm{crit}$ of void formation at temperatures between $-30$ and $90\,\mathrm{\degree C}$. The points where the voltage deviates from linearity with current by $5\,\%$ are indicated. Per cell modification and temperature, the highest $j_\mathrm{crit}$ cell out of three sample cells is selected. The different cell modifications are schematically depicted, starting with a) $\mathrm{Na_{1.9}Al_{10.67}Li_{0.33}O_{17}}$, b) $\mathrm{Na_{3.4}Zr_2Si_{2.4}P_{0.6}O_{12}}$ and c) $\mathrm{Na_5SmSi_4O_{12}}$ SEs, followed by series d) with $\mathrm{Na_{3.4}Zr_2Si_{2.4}P_{0.6}O_{12}}$ SEs and a modified Na anode with reduced grain size and series e) with $\mathrm{Na_{3.4}Zr_2Si_{2.4}P_{0.6}O_{12}}$ and a Sn interlayer between SE and Na anode.}
    \label{fig:Measurement Data}
\end{figure}

Measurements are performed with $\mathrm{Na_{1.9}Al_{10.67}Li_{0.33}O_{17}}$, $\mathrm{Na_{3.4}Zr_2Si_{2.4}P_{0.6}O_{12}}$, and $\mathrm{Na_5SmSi_4O_{12}}$ SEs. For each of the three SEs and each temperature, three cells are tested ($15$ cells per SE). To reduce measurement error, only the cell exhibiting the highest $j_\mathrm{crit}$ is selected at each temperature. This choice is justified because systematic errors (e.g., incomplete contact or interfacial contamination) bias $j_\mathrm{crit}$ to lower values, whereas no systematic errors of comparable magnitude are expected to falsely increase $j_\mathrm{crit}$. Accordingly, the results are not expected to be normal distributed, and the arithmetic mean would be biased downward relative to the real value. For the selected cells, $U(t)$ and $j(t)$ are shown in \textbf{Figure \ref{fig:Measurement Data}}a–c, with linear behavior indicated by dashed lines. The critical point at which nonlinearity is identified is marked at each temperature. A summary of all measurements is added in the Supporting Information S3. 

\begin{figure}[h]
\centering
\includegraphics[width=\linewidth]{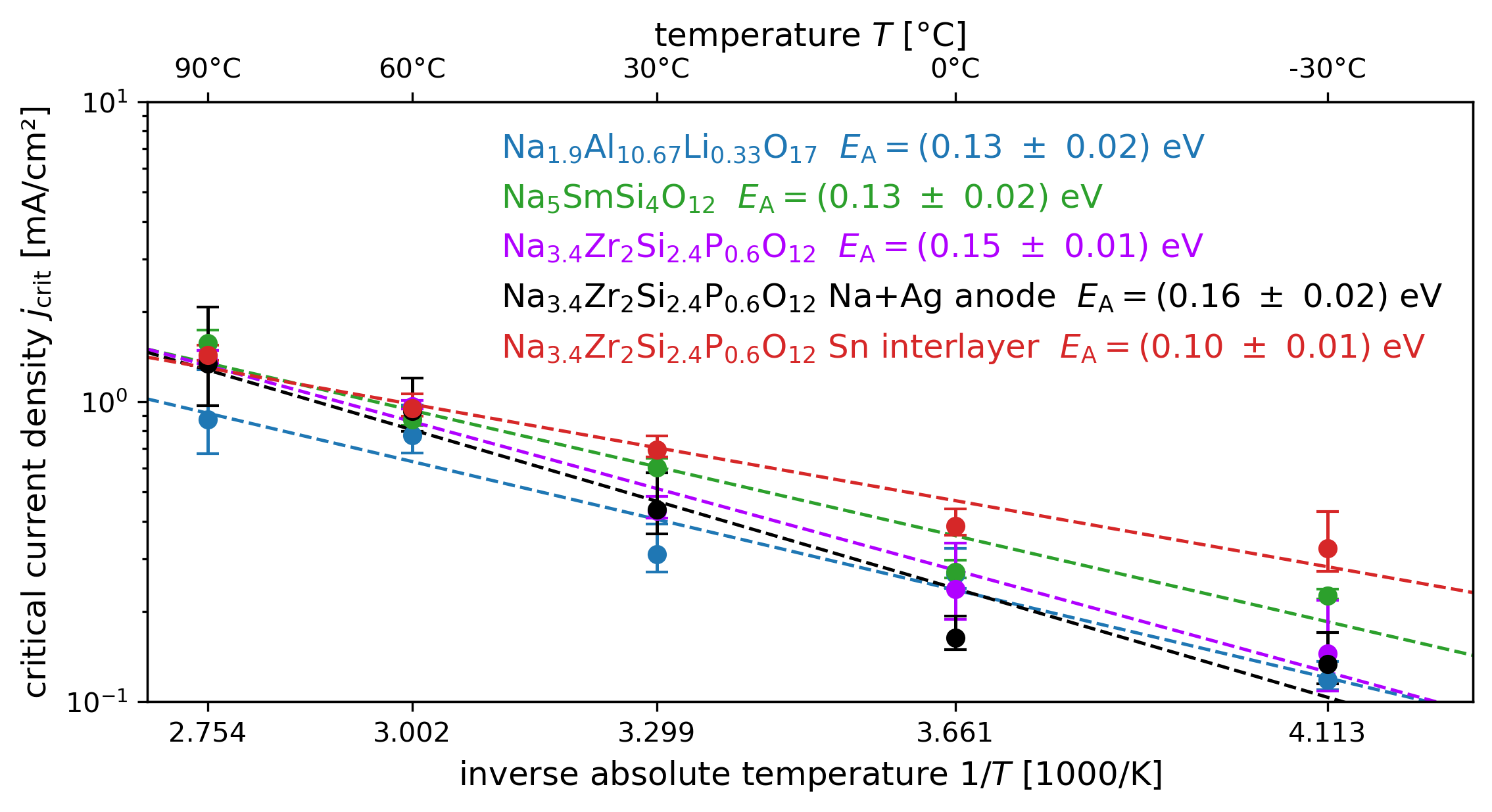}
\caption{Determination of activation energies $E_\mathrm{A}$ from Arrhenius fits of the critical current densities determined in Figure \ref{fig:Measurement Data}, plotted against inverse temperature.}
\label{fig:Measurement Arrhenius}
\end{figure}

In \textbf{Figure \ref{fig:Measurement Arrhenius}}, the extracted $j_\mathrm{crit}$ values are plotted against inverse absolute temperature $1/T$ and fitted with an Arrhenius relation, yielding $E_\mathrm{A}=(0.13\pm0.02)\,\mathrm{eV}$ for $\mathrm{Na_{1.9}Al_{10.67}Li_{0.33}O_{17}}$, $(0.15\pm0.02)\,\mathrm{eV}$ for $\mathrm{Na_{3.4}Zr_2Si_{2.4}P_{0.6}O_{12}}$, and $(0.13\pm0.02)\,\mathrm{eV}$ for $\mathrm{Na_5SmSi_4O_{12}}$. These values do not differ significantly, indicating the same limiting mechanism across all three SEs. The determined activation energies $E_\mathrm{A}$ are independent of cell geometry and therefore of substantially greater scientific value than the absolute magnitudes of the measured $j_\mathrm{crit}(T)$ values for the three SEs (as discussed above). It should be noted however, that the measured values of $j_\mathrm{crit}(T)$ tend to be smaller than the critical current densities of sodium reduction which are reported in the literature \cite{VIRKAR.1979, Ning.2023, Geng.2023, Liu.2020, Yu.2021, Lowack.2025}. Between sodium dendrite penetration during reduction and vacancy accumulation during oxidation, this indicates the vacancy accumulation to be the rate limiting mechanism of the Na-metal electrode in ceramic-SSBs (although due to feedback, the cell might still fail due to dendrite penetration in the end).

Given the activation energy $E_\mathrm{m}^\mathrm{lat}= (0.053\pm0.001)\,\mathrm{eV}$ for vacancy migration in the Na lattice \cite{Ma.2019, VSchott.2000, Ullmaier1991}, the measured $E_\mathrm{A}$ values do not satisfy equation \ref{eq:crit1_activation_energy}. A bottleneck due to vacancy diffusion within bulk Na is therefore unlikely.

\begin{figure}[h]
    \centering
    \includegraphics[width=0.5\linewidth]{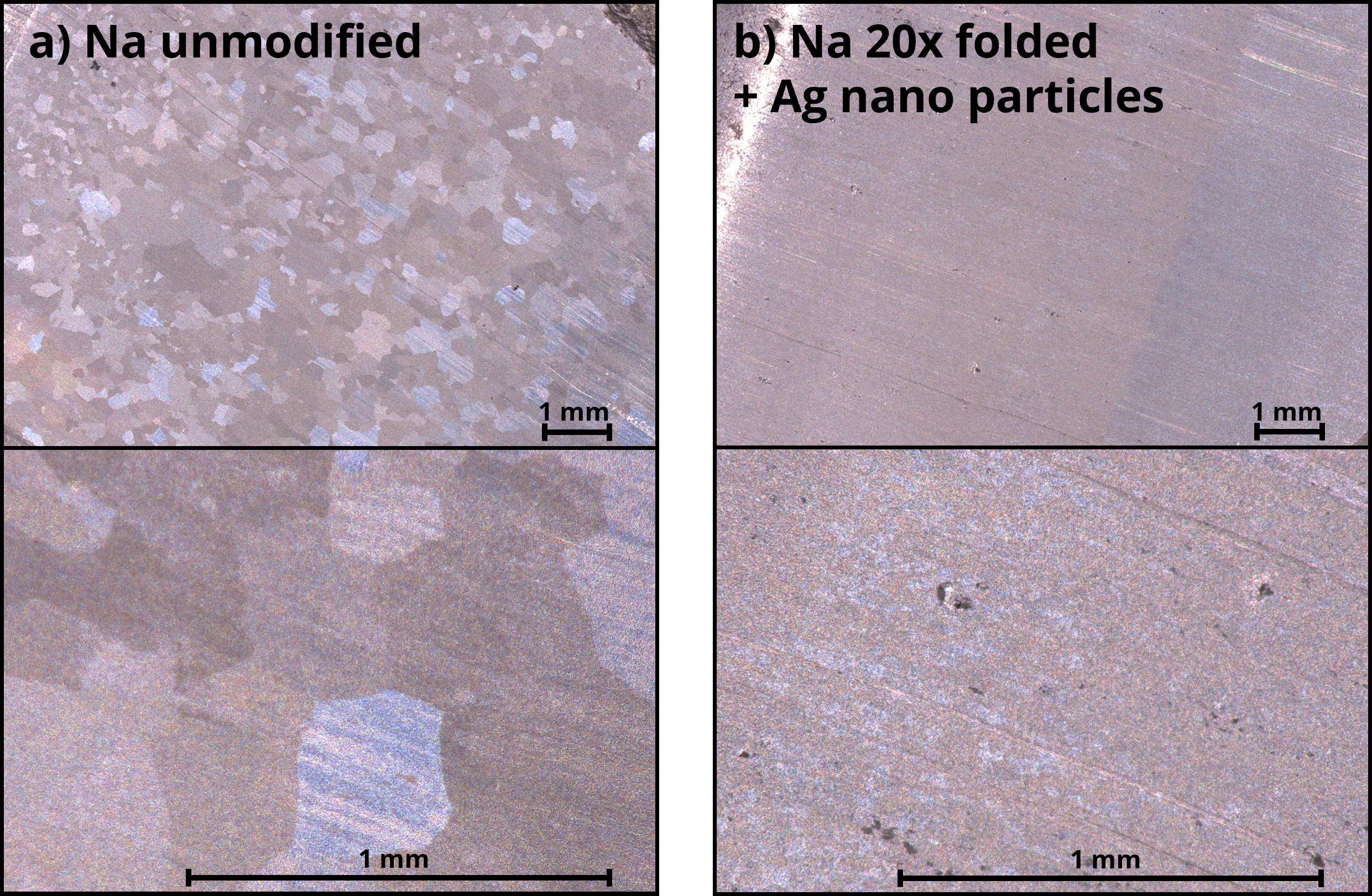}
    \caption{Optical microscopy of etched Na microstructure in two magnifications, a) unaltered microstructure, b) microstructure with reduced grain sizes after folding Ag nanoparticles ($2\,\mathrm{volume\,\%}$) $20$ times into Na.}
    \label{fig:Na microstructure}
\end{figure}

To further test this conclusion, the measurement series is repeated for $\mathrm{Na_{3.4}Zr_2Si_{2.4}P_{0.6}O_{12}}$ with an altered Na metal anode microstructure exhibiting significantly reduced grain size. Results are plotted and fitted in Figure \ref{fig:Measurement Data}d and Figure \ref{fig:Measurement Arrhenius}. The microstructure of the unmodified Na metal used for anode and cathode elsewhere in this study is shown in \textbf{Figure \ref{fig:Na microstructure}}a, while the modified morphology is shown in Figure \ref{fig:Na microstructure}b. The modified anode is fabricated by folding Na foil $20$ times, with silver (Ag) nanoparticles applied after each folding step, resulting in an Ag content of $2\,\mathrm{vol.\%}$ and a total of $1{,}048{,}576$ layers, each with a theoretical thickness below the Na lattice constant. The folding likely amorphizes the metal, and the Ag nanoparticles inhibit subsequent Na grain growth via Zener pinning. Ag was selected as it will not react which the SE. The volume percentage was chosen low enough, so it does not significantly decrease the kinetics of charge-transfer at the interface via geometric effects. Without nanoparticles, the Na morphology exhibits little change to Figure \ref{fig:Na microstructure}a after folding and $24\,\mathrm{h}$ rest at room temperature, consistent with Na’s low recrystallization temperature. The Arrhenius fit of the Ag treated anode yields $E_\mathrm{A}=(0.16\pm0.02)\,\mathrm{eV}$. This does not differ significantly from the result of  $(0.15\pm0.02)\,\mathrm{eV}$ for the unmodified Na anode with $\mathrm{Na_{3.4}Zr_2Si_{2.4}P_{0.6}O_{12}}$. Thus, the Na-anode microstructure has no measurable influence on the activation energy of $j_\mathrm{crit}(T)$. Together with the mismatch between the measured $E_\mathrm{A}$ and equation \ref{eq:crit1_activation_energy}, this supports the conclusion that equation \ref{eq:crit1} is not violated in the experiments. 

Consequently, vacancy transport should be limited by the second critical-current condition (equation \ref{eq:crit2}), assuming the model describes this transport with sufficient accuracy. Testing this hypothesis directly is challenging because modifying the Na/SE interfacial tension (and thus $E_\mathrm{f}^\mathrm{int}$ in equation \ref{eq:crit2}) without substantially altering interfacial ion-transfer kinetics is nontrivial. Varying the SE alone does not provide discrimination: as discussed above, $\mathrm{Na_{1.9}Al_{10.67}Li_{0.33}O_{17}}$, $\mathrm{Na_{3.4}Zr_2Si_{2.4}P_{0.6}O_{12}}$, and $\mathrm{Na_5SmSi_4O_{12}}$ yield similar $E_\mathrm{A}$. This would only falsify the hypothesis if the ceramics had markedly different interfacial tensions towards Na. However, this is not the case as polished samples of all three materials exhibit comparable contact angles of $(124\pm5)\,\degree$ (Figure S1) towards molten Na at $110\,\degree\mathrm{C}$. Consequently, no scientific insight into the validity of the hypothesis stated in the second critical current condition (equation \ref{eq:crit2}) is gained by a SE variation.

Some insight in this respect is gained, though, by modifying the anode side of the $\mathrm{Na_{3.4}Zr_2Si_{2.4}P_{0.6}O_{12}}$ SE with a $35$-atomic-layers (or $10\,\mathrm{nm}$) sputtered tin (Sn) film. The standard Na anode (morphology in Figure \ref{fig:Na microstructure}a is applied thereafter on top of the Sn layer. The Sn subsequently alloys with Na during a holding step at $97\,\degree\mathrm{C}$ for $20\,\mathrm{h}$ (Supporting Information S4). Here, an approximatly $55\,\mathrm{nm}$ thick Na-rich alloy phase is formed \cite{Baggetto.2013, Stratford.2017}.

\begin{figure}[h]
    \centering
    \includegraphics[width=0.6\linewidth]{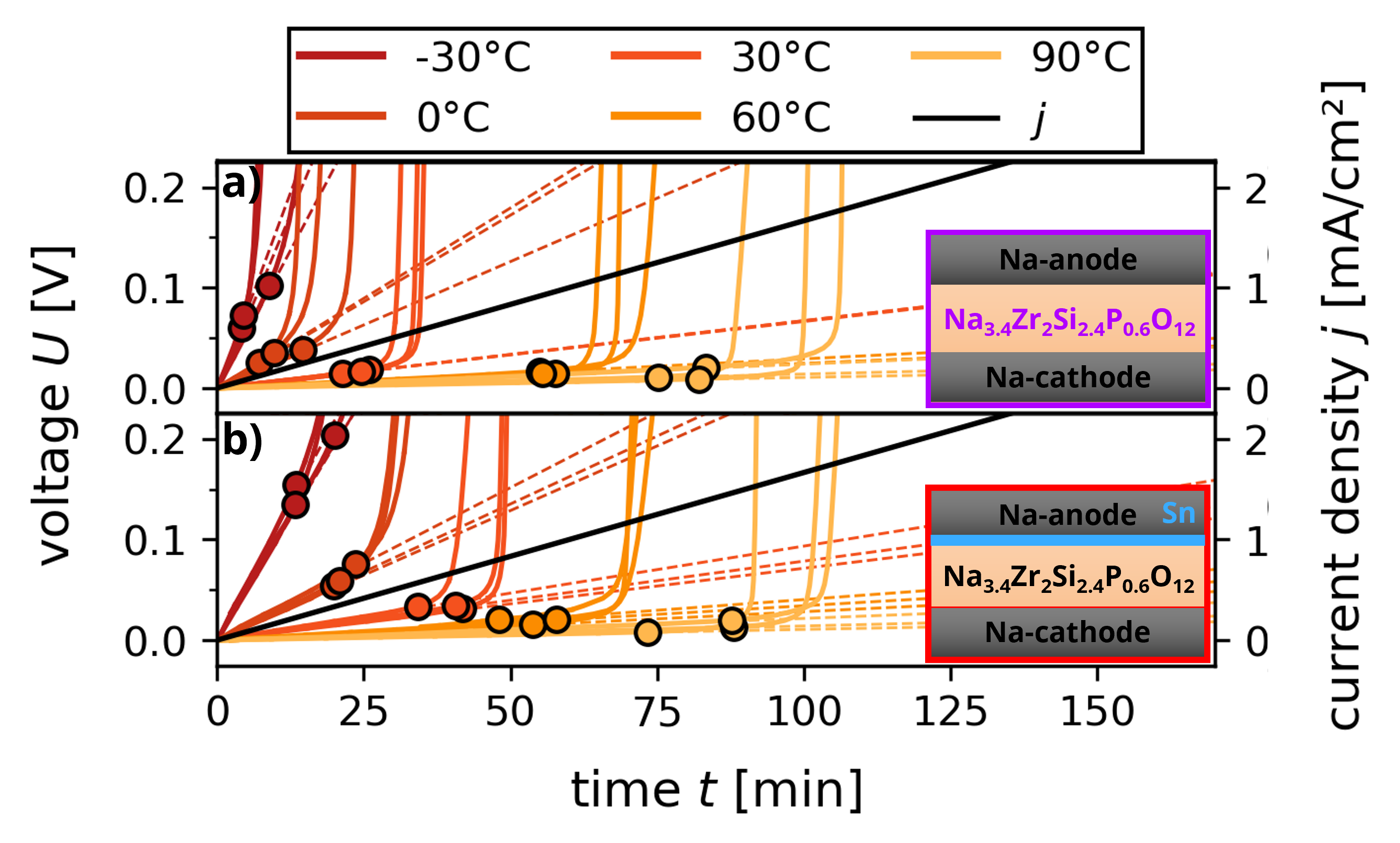}
    \caption{Determination of the critical current density $j_\mathrm{crit}$ of void formation at temperatures between $-30$ and $90\,\degree\mathrm{C}$ for a) cells with $\mathrm{Na_{3.4}Zr_2Si_{2.4}P_{0.6}O_{12}}$ SEs and unmodified Na anode and b) with $\mathrm{Na_{3.4}Zr_2Si_{2.4}P_{0.6}O_{12}}$ SE, and Sn-Na alloy interlayer in contact to the sodium anode. Three cells per modification and temperature step. The points where the voltage becomes non-linear to the current by $5\,\%$ are indicated. For $T\leq30\,\degree\mathrm{C}$, the interlayer increases $j_\mathrm{crit}$ in all cases.}
    \label{fig:Interlayer Deviation}
\end{figure}

The temperature-dependent $j_\mathrm{crit}(T)$ is measured in $15$ such test cells, with results in Figure \ref{fig:Measurement Data}e and Figure \ref{fig:Measurement Arrhenius}. An activation energy of $E_\mathrm{A}=(0.10\pm0.01)\,\mathrm{eV}$ is obtained. Although the change is small, it is statistically meaningful: at $-30$, $0$, and $30\,\degree\mathrm{C}$, all three cells at each temperature exhibit higher $j_\mathrm{crit}$ than their three counterparts using unmodified Na anodes on $\mathrm{Na_{3.4}Zr_2Si_{2.4}P_{0.6}O_{12}}$ from the same batch and preparation, as depicted in \textbf{Figure \ref{fig:Interlayer Deviation}}. The probability of this arising by chance is $1/512$ (i.e., $(1/2)^9$), indicating a significant effect of the interlayer on $j_\mathrm{crit}(T)$.

Na vacancies must traverse the interlayer into the bulk Na metal, as the alloy layer alone cannot supply sufficient Na to sustain the reaction until $j_\mathrm{crit}(T)$ is reached. Vacancy diffusion within the alloy could in principle limit transport. However, this can be excluded since the vacancy migration barrier in the alloy would be larger than in bcc Na (following the standard argumentation for vacancy diffusion in dilute alloys \cite{Wang.2025}). It therefore cannot explain a decrease in $E_\mathrm{A}$. The remaining explanation is an effect of the interlayer on the interfacial thermodynamics and thereby the vacancy transport at the Na/SE interface. Most likely, the tension at the SnNa/SE interface is significantly reduced in comparison to the Na/SE interface, reducing the energy barrier in equation \ref{eq:crit2}. 

Combined with the finding that bulk Na vacancy diffusion is not rate-limiting, this identifies the Na/SE interface as the vacancy transport bottleneck.

\section{Conclusion}
\label{conclusion}
One of the central challenges of Na-SSBs is the accumulation of atomic vacancies at the interface between ceramic SEs and Na metal anodes during discharge, i.e. Na oxidation. This effect was probed in solid-state cells with a Na metal anode and cathode. With a current-ramp protocol at $1\,\mathrm{mAcm^{-2}h^{-1}}$, a critical current density $j_\mathrm{crit}$ was detected above which vacancies accumulate rapidly enough to delaminate the Na metal, leading to irreversible cell failure. By varying the temperature of the $j_\mathrm{crit}$ measurement between $-30$ and $90\,\degree \mathrm{C}$, the activation energy $E_\mathrm{A}$ of the limiting transport mechanism was experimentally determined. For three ceramic SEs ($\mathrm{Na_{1.9}Al_{10.67}Li_{0.33}O_{17}}$, $\mathrm{Na_{3.4}Zr_2Si_{2.4}P_{0.6}O_{12}}$, $\mathrm{Na_5SmSi_4O_{12}}$), the measured activation energies cluster at $(0.13$ – $0.15)\,\mathrm{eV}$. This value is significantly higher than the activation energy $\leq0.053\,\mathrm{eV}$ expected for transport limitations due to vacancy diffusion in Na metal. A vacancy transport inside the Na metal thus cannot be the bottleneck. Especially because the measured temperature dependence and thus activation energy is insensitive to significant variations in the Na anode microstructure. In contrast, introducing a thin Sn–Na alloy interlayer significantly increases $j_\mathrm{crit}$ at low temperatures $T\leq 30\,\degree \mathrm{C}$ and shifts $E_\mathrm{A}$ to $(0.10\pm0.01)\,\mathrm{eV}$. This indicates that interfacial thermodynamics govern vacancy transfer and that the Na/SE interface constitutes the transport bottleneck. 

These findings identify the reduction of the interfacial vacancy-excitation barrier as a key design target for suppressing void nucleation and delamination during Na-SSB discharge. Sodiophilic, Na‑conducting interlayers and surface chemistries that reduce interfacial tension should be prioritized over microstructural refinement of the Na metal anode. For Na‑SSBs, this experimentally concludes an ongoing theoretical debate \cite{Seymour.2021,Yoon.2023}. The same result is likely applicable to Li-SSBs, due to the very similar values of the migration barrier $E^\mathrm{lat}_\mathrm{m}$ for vacancies in Li-metal \cite{Ma.2019, VSchott.2000, Ullmaier1991}.  However, the experiments should be repeated in the future for the most relevant ceramic Li-SEs (i.e. $\mathrm{Li_7La_3Zr_2O_{12}}$).

To identify suitable interlayers, interfacial vacancy formation energy $E_\mathrm{f}^{\mathrm{int}}$ across relevant interfaces (e.g., via ab initio calculations and advanced interfacial tension measurements for solid/solid interfaces), and wetting metrics may be correlated with the activation energy of $j_\mathrm{crit}(T)$. By engineering the interfacial thermodynamics, Na-SSBs could be driven closer to stable, high-rate operation and technologically very challenging high stack pressures above the yield strength of Na metal could be avoided.

\section{Experimental Section}
\textbf{Electrolyte Preparation}: The synthesis of the SE powders is discussed in detail in \cite{Anton.2025} for $\mathrm{Na_5SmSi_4O_{12}}$, in \cite{Ma.2019b,Dashjav.2025} for $\mathrm{Na_{3.4}Zr_2Si_{2.4}P_{0.6}O_{12}}$ and in \cite{Fertig.2022b} for $\mathrm{Na_{1.9}Al_{10.67}Li_{0.33}O_{17}}$. After synthesis, $\mathrm{Na_5SmSi_4O_{12}}$ and \\$\mathrm{Na_{3.4}Zr_2Si_{2.4}P_{0.6}O_{12}}$ powders were calcinated ($\mathrm{Na_5SmSi_4O_{12}}$: $950\,\degree\mathrm{C}$/$1\,\mathrm{h}$; $\mathrm{Na_{3.4}Zr_2Si_{2.4}P_{0.6}O_{12}}$: $1300\,\degree\mathrm{C}$/$0.5\,\mathrm{h}$). Solid electrolyte pellets were pressed by uniaxially compressing the powders. Green pellets were sintered ($\mathrm{Na_5SmSi_4O_{12}}$: $1050\,\degree\mathrm{C}$/$1\,\mathrm{h}$; $\mathrm{Na_{3.4}Zr_2Si_{2.4}P_{0.6}O_{12}}$: $1350\,\degree\mathrm{C}$/$0.5\,\mathrm{h}$; $\mathrm{Na_{1.9}Al_{10.67}Li_{0.33}O_{17}}$: $1600\,\degree\mathrm{C}$/$0.5\,\mathrm{h}$). Sintered samples were sanded in Ar atmosphere with P360 grit SiC sanding paper to achieve parallel surfaces and a thickness of $1$ to $1.1\,\mathrm{mm}$. $\mathrm{Na_5SmSi_4O_{12}}$ and $\mathrm{Na_{1.9}Al_{10.67}Li_{0.33}O_{17}}$ samples were annealed at 700 °C for 2 h in Ar atmosphere to remove potential surface layers due to reaction with air humidity (muffle furnace inside the glovebox). For the three electrode measurements samples at an increased thickness of $3\,\mathrm{mm}$ were prepared following the same procedure.

\textbf{Electrode Preparation}: To contact Na metal electrodes, SE samples were sandwiched between two Na foils and pressed at $10\,\mathrm{MPa}$ and $97\,\degree\mathrm{C}$ in a heated press inside the glovebox. This pressure ensures intimate contact and low initial interface resistance \cite{Ortmann.2023, Lowack.2025c}. Na foil was fabricated by compressing Na metal (Merck, purity $\geq99.0\,\%$) to a thickness of approximately $500\,\mathrm{\mu m}$ in the same press at room temperature. To fabricate Na foil at reduced grain size, such Na foil was coated with Ag nanoparticles (iolitec nanomaterials, $50-60\,\mathrm{nm}$), folded and subsequently again compressed to $500\,\mathrm{\mu m}$. This was repeated for a total of $20$ times, alternating folding direction. To alter the interface with an interlayer, 35 atomic layers of Sn were sputter deposited on the SE with a magnetron AC-sputter deposition source (OPTIvap Series 3G Magnetron-Sputtersource Gencoa) inside a vacuum chamber connected to the glovebox. The deposition was carried out in Ar atmosphere at $6\,\mathrm{\mu bar}$ pressure, a plasma power of $50\,\mathrm{W}$ and with $2\,\mathrm{inch}$ Sn sputter target ($99.99\,\%$ purity, Kurt J. Lesker Company). Layer thickness was controlled via deposition time and sputter deposition rate. The latter was calibrated by depositing a thicker layer for a longer time onto three test wafers of known weight. From the weight of the deposited layer and the deposition time, the deposition rate was calculated. The results were verified by thickness measurement via an X-ray fluorescence meter (FISCHERSCOPE® X-RAY XDV®-SDD, Helmut Fischer GmbH). Once the anode side of the SE was Sn coated, Na foil was applied to it as before. To ensure complete alloying of the interlayer, the cell stack was kept at $97\,\degree\mathrm{C}$ for $20\,\mathrm{h}$ (an experimental proof that this ensures alloying can be found in the Supporting Information S4). The $3\,\mathrm{mm}$ thick samples were contacted to a dot of Na metal on their side, after top and bottom electrodes had been contacted to the pellet.

\textbf{Visualisation of Na Microstructure}:The folded Na foil were held at $90\,\degree\mathrm{C}$ for approximately $6\,\mathrm{h}$ prior to imaging to give enough time for complete recrystallization. Na samples were cut using a modified doctor blade in Ar atmosphere to prepare a smooth surface. Subsequently the samples were kept uncovered in the glovebox for more than $24\,\mathrm{h}$. The remaining contamination in the glovebox atmosphere formed a thin oxide layer on the Na surface, revealing the microstructure. Imaging was conducted with an optical microscope located in the same glovebox (Keyence Digital Microscope). 

\textbf{Cell Preparation and Measurement}: Cell stacks with two electrodes were hermetically sealed in ECC-Ref (EL-CELL GmbH) cell housings using a modified spring to reduce stack pressure to approximately $10\,\mathrm{kPa}$.
In this range, the effect of stress-driven Na creep on interfacial stability is neglectable (yield strength of Na is roughly $200\,\mathrm{kPa}$ \cite{Wang.2020}, and pressures above $1\,\mathrm{MPa}$ are required to achieve intimate contact between Na electrodes and ceramic ion conductors \cite{Ortmann.2023, Lowack.2025c}). Higher stresses might be present at sharp surface artifacts, where Na creep might locally contribute to the mechanism. However, this effect would effectively lower the interfacial area on which vacancies first start accumulating by an insignificant margin (e.g. the tips of the surface artifacts). Hence, a significant effect on the measured values of $j_\mathrm{crit}$ is not likely.\\ Cells were transferred to a temperature controlled chamber and connected to a VMP3 potentiostat (BioLogic).\\ Cell stacks with three electrodes were connected inside of the glovebox to the potentiostat, using three needles and no external pressure. These measurements were conducted at the glovebox temperature of $23\,\degree\mathrm{C}$.

\textbf{Data Evaluation}: To identify the onset of nonlinearity in Figure \ref{fig:Measurement Data} and \ref{fig:Interlayer Deviation}, the recorded voltage trace $U(t)$ is first smoothed with a centered moving-average window of $30\,\mathrm{s}$ to suppress noise. The numerical second derivative $\frac{d^{2}U}{dt^{2}}$ is then computed, and $t_{1}$ is defined as the first time at which $\frac{d^{2}U}{dt^{2}} > 0.2\,\mathrm{V}\mathrm{s}^{-2}$, an empirically chosen threshold marking the rapid-increase region. A linear fit to $U(t)$ is performed over $[t_{0},\, 0.7\,t_{1}]$ with $t_{0}=30\,\mathrm{s}$; the dashed lines in Figure \ref{fig:Measurement Data} and \ref{fig:Interlayer Deviation} show these fits. The onset of nonlinearity, $t_{\mathrm{dev}}$, is defined as the first time at which the smoothed voltage deviates by more than $5\%$ from the fit, i.e.:
\begin{equation}
    \frac{\lvert U(t)-U_{\mathrm{fit}}(t)\rvert}{\lvert U_{\mathrm{fit}}(t)\rvert} > 0.05.
\end{equation}
The error of the data points in Figure \ref{fig:Measurement Arrhenius} is not symmetrically (as elaborated in the discussion). To gain a measure of experimental uncertainty, for the ordered values $j_\mathrm{crit,1}\leq j_\mathrm{crit,2}\leq j_\mathrm{crit,3}$ of the three cells per temperature and SE chemistry the uncertainty towards higher values is calculated as
\begin{equation}
    \Delta j^\mathrm{upper}_\mathrm{crit}=\frac{(j_\mathrm{crit,3}-j_\mathrm{crit,1})+(j_\mathrm{crit,3}-j_\mathrm{crit,2})}{2}.
\end{equation}
Conservatively, in figure \ref{fig:Measurement Arrhenius} half this value is used as uncertainty towards lower values of $j_\mathrm{crit}$.\\ For lower thresholds than $5\,\%$, other effects, such as nonlinear behavior of the Na cathode, could not be excluded as a source of deceptively early $j_\mathrm{crit}$ detection. For higher thresholds than $5\,\%$ (e.g. $100\,\%$, corresponding to the steep resistance increase during total contact loss), very similar activation energies between $0.1$ and $0.2\,\mathrm{eV}$ are determined. Hence, the conclusion of the study does not depend on the exact choice of the threshold.

\textbf{Use of Artificial Intelligence}: OpenAI’s GPT‑5 (via the FhGenie distribution) was used during manuscript writing for English‑language editing to improve clarity and grammar. The authors take full responsibility for the content.

\medskip
\textbf{Acknowledgments} \par
This study was supported by the State of Saxony within the M.ERA-Net 2023 framework under the project “Keramisches Anodenmaterial für definierte Natriumabscheidung” (Na-CerAnode, project11401), carried out at TU Dresden and Fraunhofer IKTS.\\\\The authors thank Jochen Rohrer and Karsten Albe for insightful input and discussion on the topic.\\\\Special thanks are extended to Marie-Theres Gerhards and Frank Tietz from Forschungszentrum Jülich for providing their $\mathrm{Na_{3.4}Zr_2Si_{2.4}P_{0.6}O_{12}}$ powder.

\medskip
\textbf{Conflict of Interest} \par
The authors declare no conflicts of interest.

\medskip
\textbf{Author Contributions} \par
A. Lowack designed the experiments, developed the theory, fabricated the model cells, conducted the experiments, evaluated the data and wrote the original draft of the manuscript. R. Anton, B. Xue and A. Lowack prepared the solid electrolyte samples. All authors contributed to the discussion and the final manuscript.

\medskip
\textbf{Data Availability Statement} \par
The data that support the findings of this study are available from the corresponding author upon reasonable request.

\newpage
\section{Supporting Information}
\subsection{S1: Contact angle measurements}\label{S1}
\begin{figure}[h]
    \centering
    \includegraphics[width=\linewidth]{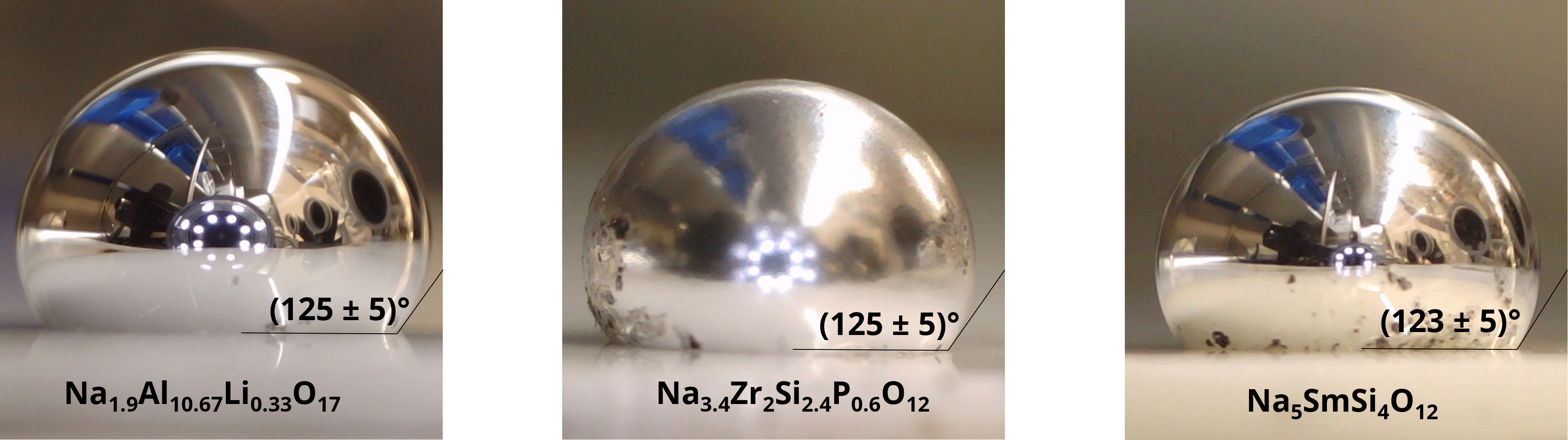}
    \caption{Contact angle measurements between different oxide ceramic solid electrolytes and molten Na at $110\,\degree\mathrm{C}$.}
    \label{fig:S1}
\end{figure}

To conduct contact angle measurements, solid electrolyte (SE) samples were prepared as discussed in the Experimental Section and polished with gradually finder SiC sanding paper up to a grit of P4000. Samples were placed on a heated plate inside an agron-filled glovebox. Temperature was adjusted until the polished SE surface measured $110\,\degree\mathrm{C}$, measured with an optical thermometer. Sodium (Na) was heated in a crucible to $150\,\degree\mathrm{C}$ inside the same glovebox. Na drops were transferred from the crucible to the SE using a gas pipette with a manually capped tip. Photographs were taken using an optical microscope, as depicted in \textbf{Figure \ref{fig:S1}}. Contact angles cluster around $124\pm5\,\degree$ for $\mathrm{Na_{1.9}Al_{10.67}Li_{0.33}O_{17}}$, $\mathrm{Na_{3.4}Zr_2Si_{2.4}P_{0.6}O_{12}}$ and $\mathrm{Na_5SmSi_4O_{12}}$. Consequently, all three materials exhibit similar interfacial tension towards molten Na, following Young's equation.

\subsection{S2: Elaboration on $\Delta j_\mathrm{crit}$}\label{S2}
The time required for voids, once nucleated, to grow to a size that measurably increases the constriction resistance is estimated: Assume that the removal of approximately ten atomic layers of Na across the entire electrode area after the onset of vacancy accumulation is sufficient to form voids that significantly affect the resistance. This corresponds to a transferred charge of $Q = 0.4\,\mathrm{\mu Ah\,cm^{-2}}$.

Let $t_\mathrm{vac}$ denote the time at which vacancy accumulation is first detected, i.e., prior to the appearance of a $5\,\%$ relative nonlinearity in the anodic interface resistance. Assuming a linear current-density ramp $j(t)=Ct$, the charge passed between $t_\mathrm{vac}$ and the time at which the current density reaches $j_\mathrm{crit}$ is
\begin{equation}
Q=\int_{t_\mathrm{vac}}^{j_\mathrm{crit}/C} j(t)\,dt
= \int_{t_\mathrm{vac}}^{j_\mathrm{crit}/C} Ct\,dt
= \frac{j_\mathrm{crit}^2}{2C}-\frac{C t_\mathrm{vac}^2}{2}.
\end{equation}
Solving for $t_\mathrm{vac}$ yields
\begin{equation}
t_\mathrm{vac}=\sqrt{\frac{j_\mathrm{crit}^2}{C^2}-\frac{2Q}{C}},
\end{equation}
where $C=1\,\mathrm{mA\,cm^{-2}\,h^{-1}}$.

The corresponding experimental overestimation of $j_\mathrm{crit}$, defined as
\begin{equation}
\Delta j_\mathrm{crit}=j_\mathrm{crit}-C t_\mathrm{vac},
\end{equation}
follows as
\begin{equation}
\Delta j_\mathrm{crit}
= j_\mathrm{crit}-\sqrt{j_\mathrm{crit}^2-2QC}.
\end{equation}
Using $Q=0.4\,\mathrm{\mu Ah\,cm^{-2}}$ and adopting a conservative lower bound of $j_\mathrm{crit}\leq 0.1\,\mathrm{mA\,cm^{-2}}$ (smaller than any value extracted from the Arrhenius plots in Figure 4 of the paper), this computes to
\begin{equation}
\Delta j_\mathrm{crit} \leq 0.004\,\mathrm{mA\,cm^{-2}}.
\end{equation}
Therefore, the resulting overestimation of $j_\mathrm{crit}$ due to delayed detection of vacancy accumulation is negligible.

\newpage
\subsection{S3: All measurements of vacancy accumulations}\label{S3}
\begin{figure}[h]
    \centering
    \includegraphics[width=0.6\linewidth]{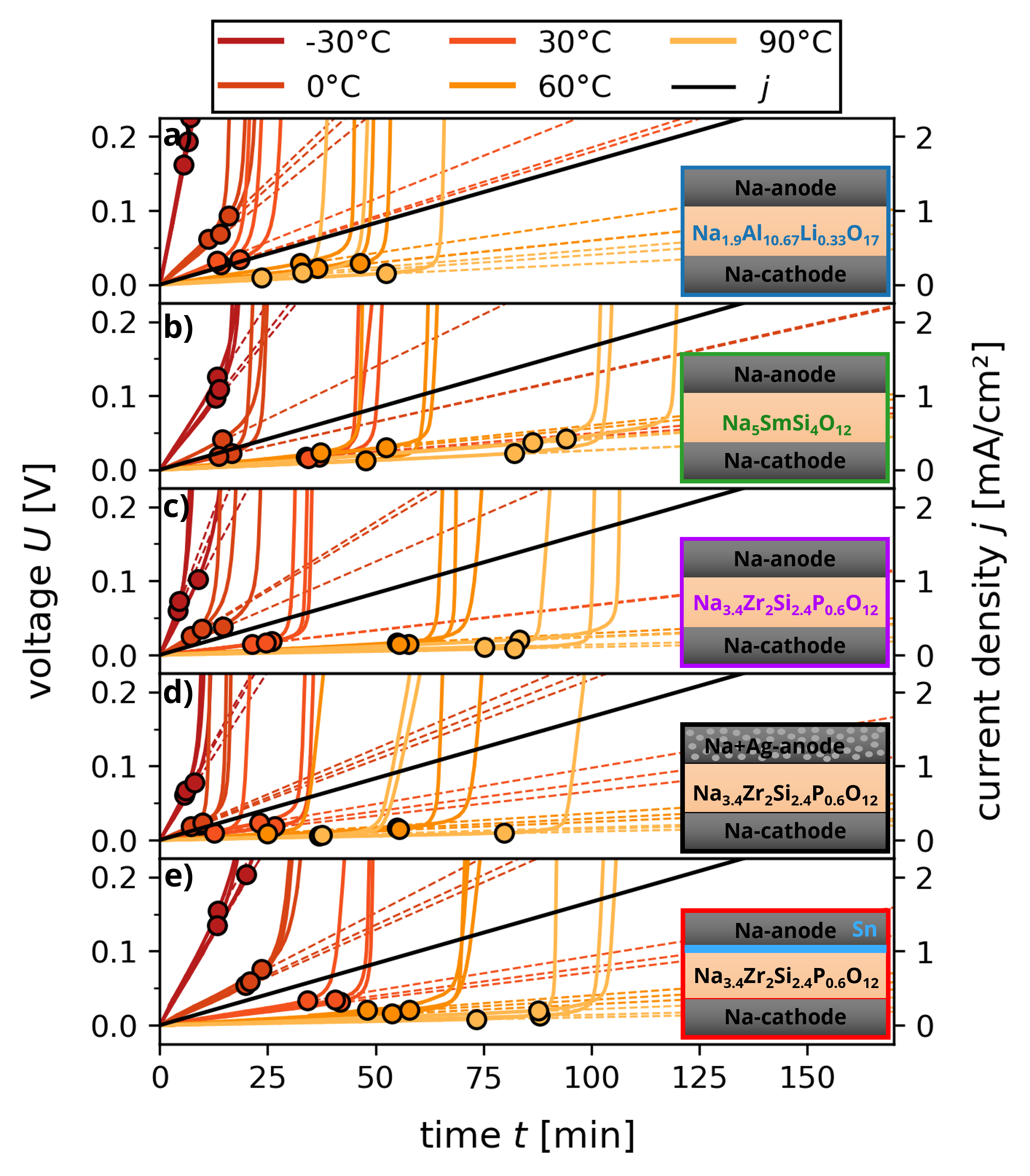}
    \caption{All measurements for the determination of the critical current density $j_\mathrm{crit}$ of void formation at temperatures between $-30$ and $90\,\mathrm{\degree C}$. The points where the voltage deviates from linearity with current by $5\,\%$ are indicated. The different cell modifications are schematically depicted, starting with a) $\mathrm{Na_{1.9}Al_{10.67}Li_{0.33}O_{17}}$, b) $\mathrm{Na_{3.4}Zr_2Si_{2.4}P_{0.6}O_{12}}$ and c) $\mathrm{Na_5SmSi_4O_{12}}$ SEs. In series d) with $\mathrm{Na_{3.4}Zr_2Si_{2.4}P_{0.6}O_{12}}$ SEs and a modified Na anode with reduced grain size. In series e) with $\mathrm{Na_{3.4}Zr_2Si_{2.4}P_{0.6}O_{12}}$ and a Sn interlayer between SE and Na anode.}
    \label{fig:S2}
\end{figure}
\newpage

\subsection{S4: Sodium-tin alloying demonstration}\label{S4}
To demonstrate the alloying of tin (Sn) with Na, a $\mathrm{Na_{3.4}Zr_2Si_{2.4}P_{0.6}O_{12}}$ sample with a diameter of $15\,\mathrm{mm}$ was prepared as described in the Experimental Section and coated with approximately $35$ atomic layers of Sn. The sample was subsequently heated to $97\,^{\circ}\mathrm{C}$, and a smaller piece of Na foil (diameter $\sim 5.5\,\mathrm{mm}$) was pressed into the center of the sample.

\begin{figure}[h]
    \centering
    \includegraphics[width=0.5\linewidth]{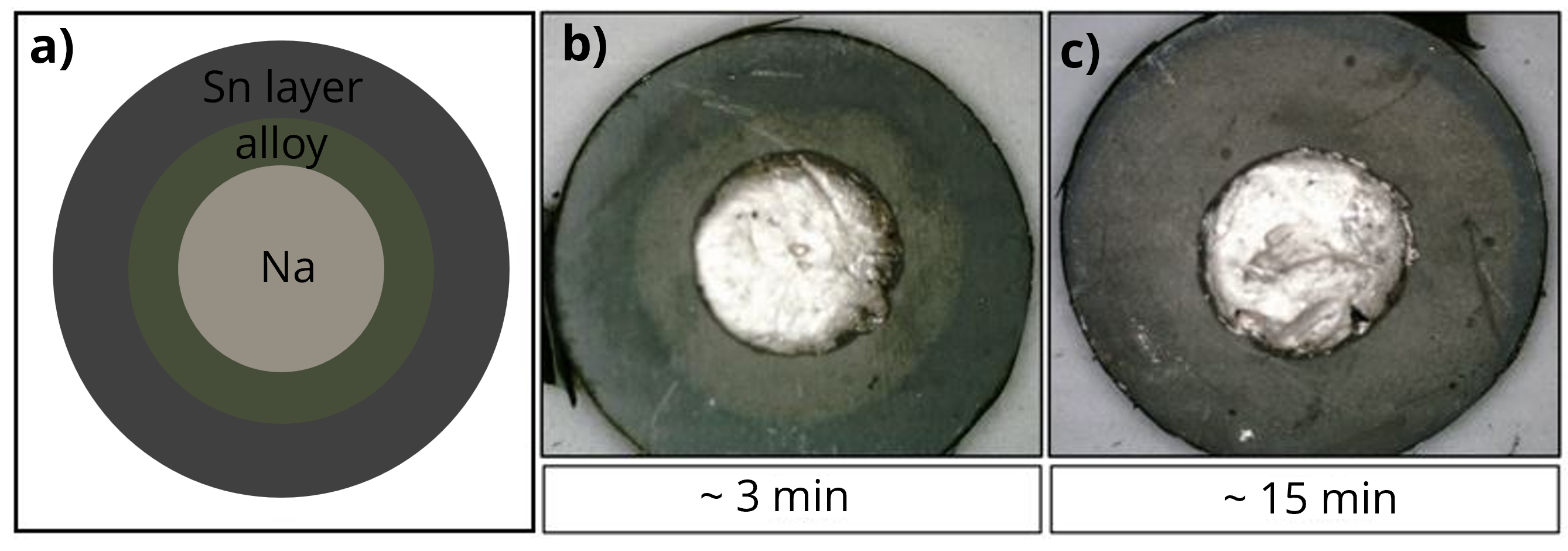}
    \caption{Optically visible alloying of solid Na on $35\,\mathrm{layer}$ Sn thin film on $\mathrm{Na_{3.4}Zr_2Si_{2.4}P_{0.6}O_{12}}$ sample at $97\,^{\circ}\mathrm{C}$.}
    \label{fig:S3}
\end{figure}

As shown in the optical microscope images in 
\textbf{Figure \ref{fig:S3}}, a discolored circular region forms around the Na foil 
shortly after contact. This region expands over the course of several 
minutes (while kept at $97\,^{\circ}\mathrm{C}$) until it covers the entire sample area. The observed 
discoloration provides direct evidence of Sn-Na alloy formation. This 
result confirms that a similar alloyed interlayer forms when Na foil is 
applied onto the Sn-coated surface during cell assembly.

\end{document}